\documentclass[aps,prd,amssymb,cite,
amsfonts,epsf,preprintnumbers,nofootinbib,superscriptaddress]{revtex4}

\usepackage[dvips]{graphicx}
\usepackage{bm,latexsym,amsmath,amssymb,amsfonts}
\usepackage[usenames,dvipsnames]{color}
\usepackage[colorlinks=true,linkcolor=blue]{hyperref}
\usepackage{color}
\usepackage{soul}
\usepackage{epsfig}
\usepackage[mathscr]{eucal}
\usepackage{cancel}
\usepackage{mathrsfs}
\usepackage{pgf, tikz}
 
\definecolor{mypink1}{rgb}{0.858, 0.188, 0.478}
\definecolor{mypink2}{RGB}{219, 48, 122}
\definecolor{mypink3}{cmyk}{0, 0.7808, 0.4429, 0.1412}
\definecolor{mygray}{gray}{0.6}
\definecolor{pptbg}{rgb}{0.961,0.945,0.863}

\newcommand{\be}[1]{\begin{equation} \label{#1}}
\newcommand{\ee}{\end{equation}}
\newcommand{\bea}{\begin{eqnarray}}
\newcommand{\eea}{\end{eqnarray}}
\newcommand{\ba}{\begin{array}}
\newcommand{\ea}{\end{array}}
\newcommand{\nn}{\nonumber}

\newcommand{\bel}{\begin{align}}
\newcommand{\eel}{\end{align}}

\newcommand{\ve}[1]{\vec{\bm{#1}}}

\begin{document}
\title{Local temperature in general relativity}

\author{Hyeong-Chan Kim}
\email{hckim@ut.ac.kr}
\author{Youngone Lee}
\email{youngone@ut.ac.kr}
\affiliation{School of Liberal Arts and Sciences, Korea National University of Transportation, Chungju 380-702, Korea}

\begin{abstract}
We examine the Tolman temperature  by using Carter's variational formalism of thermodynamics.
We restrict our interests to fluids in thermal equilibrium that the heat does not propagate.
We show that this condition presents a general formula for the local temperature gradient.
We suggest a resolution of the recently addressed conflict in Tolman temperature when a chemical potential does not vanish. 
\end{abstract}

\maketitle
 
Two physical systems in thermal contact are said to be in thermal equilibrium if there is no net flow of thermal energy between them. 
There may exist chemical or mechanical differences between the two systems even if they are in thermal equilibrium.
It is characterized by a physical parameter, temperature. 

In considering general relativity,
Tolman~\cite{Tolman,Tolman2} discovered that there exist relativistic temperature gradients for fluids in thermal equilibrium in static spacetimes in 1930. 
The locally measured temperature $\Theta(x^i)$ is 
\be{Tolman}
\Theta(x^i) = \frac{T_0}{\sqrt{-g_{00}(x^i)}},
\ee
where $g_{00}$ is the time-time component of the metric on a static geometry, and $T_0$ is a constant.
$T_0$ represents the physical temperature at the zero gravitational potential hypersurface~\cite{Santiago:2018kds}.
In 1949, Buchdahl~\cite{Buchdahl:49} extended Tolman's result that looks identical to Tolman's.
They considered fluids in stationary spacetimes following a timelike Killing vector.
Recently, the non-uniqueness of the Killing vector and the temperature gradient was explored by Santiago and Visser~\cite{Santiago:2018lcy}.
The study is based on the gradient flow normal to a spacelike hypersurface.
The authors also argued that the temperature gradients take the same form for two distinct systems composed of different materials.
The argument is based on the universality of gravity~\cite{Santiago:2018kds}:
If the temperature depends on a specific material property, then one can make a perpetual motion machine.
This argument using the `evading a perpetuum mobile' is so powerful that it is difficult to refute.
 
On the other hand, Lima et. al.~\cite{Lima:2019brf} recently argued that the original Tolman temperature should be modified 
 for fluids in a static spacetime when its chemical potential does not vanish.
They explicitly solved the energy conservation law for a general equation of state and used
the conservation of particle number and entropy.
The discrepancy between the two arguments seems so different that it cannot be compromised easily.  

There is another approach to derive the temperature gradient by Cocke~\cite{Cocke}.
He also derived the Tolman-Oppenheimer-Volkoff equation (TOV)~\cite{Tolman3,Oppenheimer} through a maximum entropy principle which was further extended by Sorkin et. al.~\cite{Sorkin}. 
Roupas~\cite{Roupas1,Roupas2,Roupas3,Roupas4} recalculated the TOV,
 the Tolman's, and  Klein's results~\cite{Klein49} after specifying an appropriate thermodynamic ensemble. 
It is also worth mentioning that Rovelli and Smerlek~\cite{Rovelli} obtained the Tolman temperature by applying the equivalence principle to a property of thermal time. 
The quantum mechanical modification of Tolman temperature based on the trace anomaly was also proposed by Gim and Kim~\cite{YGim15}. 

To examine the above schism,
we use Carter's axiomatic approach for relativistic thermodynamics~\cite{Carter72,Carter73,Carter89,Andersson:2006nr} to derive the Tolman temperature.
The approach has the same level of generality as the Israel-Stewart theory~\cite{IS1,IS2,IS3}.
The theory generalizes Eckhart's thermodynamics~\cite{Eckart:1940aa} in the context of general relativity.
Moreover, the two theories are equivalent in the limit of linearized perturbations around a thermal equilibrium state. 
The two theories coincide as far as the causal property is concerned~\cite{Priou91}. 
The original Carter's theory was further generalized to consider viscosity and resistivity 
by Andersson and Comer~\cite{Andersson2006}. 

The original Carter's theory dealt with a general two-constituent, 
two-fluid model composed of the caloric (entropic) flow $s^a$  and the number flow $n^a$ of matter. 
There, the number flows was generalized to describe various matter flows by introducing a number flow vector for each kind of matter.

In general, the two flows move freely from each other and allow heat conduction. 
However, in thermal equilibrium, there is no heat conduction.
Thus we restrict our interest to the case that the two constituents $s^a$ and $n^a$ are parallel, $s^a\parallel n^a$. 
It is to demonstrate the essential parts of what we want to describe avoiding technical complexities due to entrainments and heat flow. 
In the previous literature~\cite{Carter89,Andersson2006,Andersson:2006nr,Andersson2011}, 
the authors had dealt with the original single-fluid model in more general forms.
We display the results which are necessary for the present work.  
Mostly, we follow the notations of Carter~\cite{Carter89}.

The variational theory begins with an unidentified-master function, $\Lambda(s,n)$, of two scalars $s$ and $n$ given by
\be{sn:sf}
s = (-s_a s^a)^{1/2}, \qquad n = (-n_a n^a)^{1/2}.
\ee
When the caloric flow $s^a$ is not parallel to the number flow $n^a$, the master function depends on another scalar, $x\equiv -s_a n^a$, additionally.
However, $x$ is not independent
 because $x=sn$ ($s^a \parallel n^a$).
The master function can be integrated to compose an action functional for $s$ and $n$,
\be{S:sf}
I = \int_{\mathcal{M}} d\mathcal{M}\, \Lambda(s,n),
\ee
where $d\mathcal{M}$ represents the four-dimensional spacetime volume form.

It is useful to introduce a unit tangent vector $u^a$ along the flow-lines of $n^a$ by 
\be{u:def}
u^a \equiv \frac{n^a}{n}, \qquad n \equiv \sqrt{ -n^a n_a}, \qquad u^au_a =-1,
\ee
where $u_a \equiv g_{ab} u^a$ and $g_{ab}$ is the metric of the manifold $\mathcal{M}$.
Then, the entropy flow can be written as
\be{s:def}
s^a = s u^a. 
\ee
The conjugate covectors to $s^a$ and $n^a$ are 
\be{conjugate:sf}
\Theta_a \equiv \Theta(s,n) u^a, \qquad
\chi_a  \equiv \chi(s,n) u^a,
\ee
where $\Theta(s,n)$ and $\chi(s,n)$ denote the temperature and the chemical potential with respect to comoving observers, which are conjugates to $s$ and $n$, respectively. 

Varying the action functional~\eqref{S:sf} with respect to the fluid path and the metric, one can find the stress tensor 
$$T_a^b= \Theta_a s^b + \chi_a n^b + \Psi g_a^{~b} = -\Lambda u_a u^b + \Psi \gamma_a^{~b},$$
where the pressure $\Psi$ and the energy density $\rho$ are
\be{Pressure}
\Psi = \Theta s + \chi n - \rho, \qquad \rho = -\Lambda.
\ee
The force densities for the caloric and the number parts, by using differential form notation, are
\bea 
 \bm{f}^0 &=& \bm{\Theta} \Gamma_s
	+ \ve{s} \cdot (\bm{d \Theta}), \label{force0}\\
\bm{f}^1 &=& \bm{\chi} \Gamma_n + \ve{n} \cdot (\bm{d \chi}).
	\label{force1}
\eea
Here $[\bm{d \Theta}]_{ab} = 2\nabla_{[a} \Theta_{b]}$ and 
	$[\bm{d\chi}]_{ab} = 2\nabla_{[a} \chi_{b]}$.
The creation rates are
\be{GG:def}
\Gamma_s \equiv \nabla\cdot \ve{s} =\dot{s}+ s\theta
, \qquad 
\Gamma_n \equiv \nabla\cdot \ve{n} = \dot{n} + n \theta,
\ee
where $\theta \equiv \nabla_a u^a$ is the expansion rate of the flow lines.
From now on, the overdot denotes the time derivative, i.e., $\dot s \equiv u^a\nabla_a s$ and $\dot n \equiv u^a \nabla_a n$, respectively\footnote{In the generalized Carter theory developed in Refs.~\cite{Celora20,Andersson2006}, the equation of motions for the forces were shown to have extra terms representing resistivity and viscosity.
}.

The divergence of the stress tensor can be decomposed into two parts, each representing the forces for the caloric and the number flows, 
\be{bianchi}
\nabla_c T^{c}_{~a} = f^0_ a + f^1_a. 
\ee
Therefore, when the fluid is isolated, the stress tensor must be conserved, $\nabla_c T^{c}_{~a} =0$, which gives 
\be{f0 f1:SF}
\bm{f}^0 = -\bm{f}^1.
\ee
The two forces work as a pair of action on the caloric part and reaction on the number part.\footnote{
This identity is satisfied even if
 the particle flux $n^a$ is no longer parallel to the entropy flux $s^a$.
It is because this  identity is the natural result of local energy-momentum conservation.
The introduction of another scalar $x\equiv -s_a n^a$ for a heat transfer does not change the result
(see Eq.~(2.26) in \cite{Carter89}).
}
It reminds us of Newton's third law.
It reads that 
the `caloric' force and the `numeric' force are equal in magnitude and opposite in direction.
When the particle flux $n^a$ is no longer parallel to the entropy flux $s^a$,
we need more equations of motion to cover the additional degrees of freedom.
An example is the Israel-Stewart version of the relativistic Cattaneo equation, $q^a=q^a(\kappa,T, T_{;c}, u^a,u^a_{;c})$, 
 Eq.~(3.35) in Ref.~\cite{Andersson2011}.
In the case of the thermal equilibrium, the equation $q^a=0$ suffices the condition.

The spatial projections of the forces on a spacelike section $\Sigma$ orthogonal to $u^a$  become
\bea \label{fperp SF}
\underline{f}^0_{~a} \equiv \gamma_a^bf^{0}_{~b} 
&=& su^c \nabla_{[c}\Theta_{a]} 
	= s\Big(\nabla_a\Theta + \frac{d}{d\tau}(\Theta u_a) \Big), \nn \\
\underline{f}^1_{~a} \equiv \gamma_a^bf^{1}_{~b} 
&=& nu^c \nabla_{[c}\chi_{a]} 
	= n\Big(\nabla_a\chi + \frac{d}{d\tau}(\chi u_a) \Big),
\eea
where we multiplied $\gamma^{b}_a = g^{b}_a+ u_a u^b $ to Eqs.~\eqref{force0} and \eqref{force1}
using the property of $\chi^a$ and $\Theta^a $  parallel to $u^a$ in Eq.~\eqref{conjugate:sf}.

\vspace{.2cm}
Starting from this formula, one can deduce the two incompatible arguments mentioned earlier. 
Let us display the two in order.
\begin{enumerate}
\item \label{item1}

When the caloric flow is inseparable from the number flow, the two forces $f^0_a $ and $f^1_a$ must cancel each other.
One can see it in Eq.~\eqref{f0 f1:SF} without knowing the internal mechanism. 
We are not interested in the precise mechanism but rather focus on the consequences in the present work.

The sum of the two forces vanishes because of the conservation law~\eqref{f0 f1:SF}, which gives
\be{de }
\left(1+\frac{\sigma \Theta}{\chi}\right) \dot u_a + \left(D_a \log \chi +\frac{\sigma \Theta}{\chi} D_a \log \Theta\right) =0 ,
\ee
where $\dot u^a = u^c\nabla_c u^a$ is the acceleration of the fluid and $D_a = \gamma_a^b \nabla_b$ is the natural covariant derivative on the spacelike section $\Sigma$.
Using Eq.~\eqref{Pressure}, the equation becomes
\be{de 2}
 \dot u_a + D_a \log \Theta
	+ \frac{n\chi}{\rho + \Psi}D_a \log \frac{\chi}{\Theta} =0.
\ee

Because $n^a$ denotes the number flow, without loss of generality, we may choose the unit vector $u^a$ to flow along the coordinate time, i.e., $ (\partial_t)^a = u^a$. 
In other words, the fluid  is irrotational, i.e., to have vanishing vorticity.
On a comoving coordinates with the number flow $n^a$, the acceleration is 
$$
 \dot u^a = u^c\nabla_c u^a=  \Gamma^{a}_{~bc} u^b u^c .
$$
Specifically, we consider the metric which does not have a shift vector:
\be{ds2:static}
ds^2 = -g_{00}(t,\vec x) dt^2 + g_{ij}(t,\vec x)  d\vec{x}^2.
\ee 

Now, the spacelike surface $\Sigma$ is described by a $t=$ constant surface.
Because $\dot u^a$ is orthogonal to $u_a$,  $u_a \dot u^a =0$, it is enough to consider only the spatial component of $\dot u^a$.  
Then, the Christoffel symbol 
$
\Gamma^{a}_{~bc} = \frac12 g^{ad} \left[ g_{bd,c} + g_{cd,b} - g_{bc,d} \right] 
$
becomes, using  $g_{00} (u^t)^2 = -1$, 
\be{dot ui}
\dot u^i = \Gamma^i_{00} (u^0)^2 = \frac {g^{ij}}{2 g_{00}}  \frac{\partial g_{00}}{\partial x^j} \quad 
\Rightarrow \quad \dot u_i =  \frac{\partial}{\partial x^i} \log (-g_{00})^{1/2} .
\ee
Using Eq.~\eqref{dot ui}, Eq.~\eqref{de 2} becomes
\be{Tolman:Lima}
\frac{\rho + \Psi}{n \chi } D_i \log \left(\sqrt{-g_{00}}\Theta \right)
	+  D_i \log \frac{\chi}{\Theta} =0.
\ee
When $\rho = -\Psi$, $\chi/\Theta $ is constant over the given spacelike surface.
When the chemical potential $\chi =0$ or $\chi/\Theta$ is spatially homogeneous, one gets the temperature in a Tolman form~\eqref{Tolman}. 
Else, the Tolman temperature is to be modified.
This possibility was reported only recently in Ref.~\cite{Lima:2019brf} for static spacetimes.
The authors explicitly pointed out that the ratio $\chi/\Theta$ does not need to be a constant over $\Sigma$ under general conditions.
In an almost complete degenerate regime ($\Theta/\Theta_F \ll 1$, $\Theta_F$ is the Fermi temperature),
one can write the chemical potential for a degenerated relativistic Fermi gas as:
$$
\chi = E_F\left[1-\frac{\pi^2}{12} \left(\frac{\Theta}{\Theta_F}\right)^2+\cdots \right] + m c^2,
$$
where $E_F = k_B \Theta_F$ is the Fermi energy. 
With this relation, it is sure that the position independence of $\chi/\Theta$ fails to hold for such fluids.
The Tolman relation does not hold anymore when Eq.~\eqref{Tolman:Lima} is satisfied.

\item \label{item2}
In the previous argument, we assume the caloric flow is inseparable from the number flow. 
This assumption is not always valid because the caloric flow flows with the heat.
As we know, heat may move independently from the matter as a form of thermal energy.
Let us review shortly the results in the literature~\cite{Priou91,Andersson2011,Olson1990}.
The authors have assumed the absence of particle creation.
In the presence of heat, the spatial part of the force $\underline{f}^0_{~a}$ has the form~\cite{Andersson2011},
\be{f decom}
\underline{f}^0_{~a} = f^{q} q_a + f^{\perp}_{~a} , 
\ee
where $f^q$ represents the force parallel to the heat $q^a$
and $f^\perp_{~a}$, orthogonal to $u^a$ and $q^a$, respectively.
The orthogonal term, in general, takes the form,
\be{f ortho}
f^\perp_{~a} = \gamma_a^c \perp_c^d\left[A_d q^2 + B\nabla_d (q^2)+ C q^b q_{d;b}\right] ,
\ee
where $A_d$, $B$, and $C$ are unspecified coefficients that are regular at $q=0$.
Here, $\perp_c^d = \delta_c^d- q_c q^d/q^2$ denotes the projection operator orthogonal to the heat direction $q^a$.
The force $f^\perp_{~a}$ is not related to the entropy production even though it is involved in the heat. 
In the absence of heat, $q=0$, we can safely set the orthogonal force to vanish
$f^{\perp}_a =0$.
Note that the parallel force to the heat also vanishes.\footnote{
This vanishing property of forces holds even for the generalized Carter's theory~\cite{Celora20}.
The caloric and the numeric forces here, $f^0_a$ and $f^1_a$, correspond to
$\bar f^x_a+\Gamma_x\bar\mu_a^x$ for $x=0,1$ in Eq.~(14) in the reference, respectively.
In thermal equilibrium, where the dissipation contributions vanish,
$\bar f^x_a+\Gamma_x\bar \mu_a^x =0$,  which confirms the vanishing property. 
Note that non-zero creation rates, $\Gamma_x\neq 0$, do not change our analysis
 because $f^0_a$ and $f^1_a$ include those rates by definition.
}
This result presents a strong constraint on the caloric flow satisfying $n^a \parallel s^a$ because it makes the spatial part of the caloric force, $\underline{f}^0_{~a}$, vanish. 
Naturally, from Eq.~\eqref{f0 f1:SF}, the number force $\underline{f}^1_{~a} $ also vanishes.

Multiplying $\gamma^{ab}$ to Eqs.~\eqref{force0} and \eqref{force1} after combining the two, we get, from $\underline{f}^0_{~a}=0=\underline{f}^1_{~a}$, the equilibrium condition for the fluid:
\be{equil}
\nabla_b \frac{\Theta}{\chi} + \left(\frac{d}{dt}\frac{\Theta}{\chi}\right) u_b  =0 \quad
\Rightarrow \quad \gamma_a^b \nabla_b \frac{\Theta}{\chi} =0.
\ee
The last equation implies that the ratio between temperature and chemical potential, $\Theta/\chi$, should be constant over $\Sigma$.
This relation is nothing but Klein's law~\cite{Klein49}.

The remaining equation in Eq.~\eqref{fperp SF} with vanishing orthogonal forces presents another equilibrium condition, 
\be{equil2}
\dot{u}^a = -D_a\log\Theta  .
\ee
This equation determines how the temperature $\Theta$ varies in space which gives the local temperature of Tolman~\cite{Tolman} for a static spacetime.
That is, using Eq.~\eqref{dot ui}, one gets $\sqrt{-g_{00}} \Theta $ is constant over $\Sigma$ for a spacetime given by the metric~\eqref{ds2:static}.
\end{enumerate}

Note that we do not assume the spacetime to be static or stationary in deriving this formula. 
The differential result follows from the unique assumption, the absence of heat.

In general, Eq.~\eqref{equil2} is not integrable.
Hence, it does not give an unambiguous temperature $\Theta$ for an arbitrary fluids in arbitrary spacelike sections.  
On a spacelike surface $\Sigma$, the temperature $\Theta$ is well-defined only when the acceleration $\dot u_a$ takes the form of a gradient of a scalar function on $\Sigma$, 
\be{du Dphi}
\dot u_a = D_a \log \phi = \gamma_a^b \nabla_b \log \phi.
\ee
Then, we get the temperature 
\be{T:phi}
 \Theta = \frac{T_0}{\phi}.
\ee

It is natural to ask whether one can define the temperature over the whole spacetime starting from the temperature on $\Sigma$.  
To answer this question, we define an exact form field 
\be{v:phi}
v_a = \nabla_a \log \phi'.
\ee
If we choose $\phi=\phi'$ on $\Sigma$, the projection of $v_a$ on the spacelike surface reproduces $\dot u_a$. 
However, $v_a \neq \dot u_a$ because $\dot u_a u^a=0$ but $v_a u^a \neq 0$ in general.
The differential form $v_a$ is closed,
$$
\nabla_{[b} v_{a]} = \nabla_{[b} \nabla_{a]} \log \phi' =0,
$$
where we used the torsion-free condition of the Einstein gravity theory.
On $\Sigma$, identifying $\phi'=\phi$ after projecting the above equation onto $\Sigma$ by using $\gamma_a^b$, we get,
$$
 \gamma_b^c\gamma_a^d \nabla_{[c} v_{d]} =D_{[b} \dot u_{a]} =0.
$$
This equation implies that the one-form field $\dot u_a \equiv \gamma_a^b v_b$ is closed on $\Sigma$, 
a necessary condition for the well-defined temperature on each $\Sigma$.
When the geometry of $\Sigma$ allows the closed one form field to become an exact form, one can define the temperature uniquely from the field $v_a$. 
For example, when $\Sigma$ is simply connected, the temperature is well-defined. 
One may apply the present approach  beyond the static spacetime must be the homogeneous Robertson-Walker spacetime in cosmology.

Finally, let us return to the previously mentioned discrepancy: 
In general, a chemical potential must be an independent parameter from temperature, 
and the two may not be proportional to each other, as was stated in Ref.~\cite{Lima:2019brf}. 
When gravity is weak, the chemical potential and the temperature may not have a spatial gradient, and Eq.~\eqref{Tolman:Lima} is automatically satisfied.
However, as the gravity effect becomes measurable, 
the non-trivial relation between the chemical potential and the temperature must falsify Klein's law. 
Let us examine the derivation process of Klein's law in detail. 
In deriving the law, we used $\underline{f}^0_{~a}=0$ when heat is absent.
The absence of the orthogonal force presents the Tolman temperature. 
We use this result and the energy-momentum conservation law of the stress tensor to derive $\underline{f}^1_{~a}=0$, 
which is crucial for the Klein's law.
Therefore, the law is an indirect consequence contrary to the Tolman temperature. 
Because of this, we can avoid the law by considering multi-constituents fluid models. 
For example, let us consider three constituents models composed of $s^a$, $n^{1a}$, and $n^{2a}$.
Now, the energy-momentum conservation law presents
$$
\nabla_c T^c_{~a} = f^0_{~a} + f^1_{~a} + f^2_{~a} =0.
$$
In this case, the condition $\underline{f}^0_{~a}=0$ fails to make the orthogonal force for matter,
 $\gamma_a^b f^{k}_{~b} ~ (k=1,2) $, vanish.  
Therefore, for each number flow, Klein's law won't hold contrary to the relation $ \gamma_a^b(f^1_{~b} + f^2_{~b}) =0$.
For the case of the degenerated relativistic Fermi gas, 
other matters such as the protons and the nucleons exist 
that keep the electron to stay around.
Therefore, when we consider the force law for matter, we should take care of them.
Those matters will contribute to the orthogonal force $f^\perp_{~a}$ in Eq.~\eqref{f decom}. 

\vspace{.2cm}
In summary, we have examined Tolman temperature by using Carter's variational formalism of thermodynamics.
To deal with the situation, we have restricted our interests to fluids in thermal equilibrium.
We also suggest a resolution of the recent conflict in the Tolman temperature when a chemical potential does not vanish. 
Our results support the generality of the Tolman temperature over different kinds of matter.
We have also argued that Klein's law does not hold in general.

\section*{Acknowledgment}
This work was supported by the National Research Foundation of Korea grants funded by the Korea government NRF-2020R1A2C1009313.

\bibliographystyle{unsrt}
\bibliography{reference}

\end{document}